\documentclass[10pt,aps,preprintnumbers,prd,noshowpacs,nofootinbib,noshowkeys,floatfix,superscriptaddress]{revtex4}
\usepackage[dvips]{graphics,graphicx}
\usepackage[colorlinks=true,linktocpage=true,linkcolor=blue,citecolor=blue]{hyperref}
\usepackage[usenames,dvipsnames]{color}
\usepackage{amsmath, amssymb,oldgerm}
\usepackage{multirow}
\usepackage{xcolor}
\usepackage[normalem]{ulem}  
\usepackage{braket}
\usepackage{slashed}
\usepackage[mathscr]{eucal}
\usepackage[dvips]{graphics,graphicx}
 \usepackage{caption}
\usepackage{subcaption}

\newcommand{\n}{\nonumber}

\newcommand{\Pc}{\mathcal{P}}

\newcommand{\beq}{\begin{equation}}
\newcommand{\eeq}{\end{equation}}

\newcommand{\eqs}{Eqs.~}
\newcommand{\eq}{Eq.~}

\newcommand{\bbb}{\textcolor{black}}
\newcommand{\p}{\mathbf{p}}
\newcommand{\Z}{\mathcal{Z}}
\newcommand{\X}{\mathcal{X}}

\newcommand{\mJ}{\mathcal{J}}
\newcommand{\mT}{\mathcal{T}}
\newcommand{\T}{\mathcal{T}^\lambda}
\newcommand{\J}{\mathcal{J}^\lambda}

\begin{document}

\title{Chiral hydrodynamics of expanding systems}

\author{Nora Weickgenannt}

\affiliation{Institut de Physique Th\'eorique, Universit\'e Paris Saclay, CEA, CNRS, F-91191 Gif-sur-Yvette, France}

\author{Jean-Paul Blaizot}

\affiliation{Institut de Physique Th\'eorique, Universit\'e Paris Saclay, CEA, CNRS, F-91191 Gif-sur-Yvette, France}

\begin{abstract}
 We obtain equations of motion for the boost-invariant expansion of a system of chiral particles. Our analysis is based on the Boltzmann equation for left- and right-handed massless particles in the relaxation time approximation. We assume Bjorken symmetry, but allow for parity breaking.  We generalize the relaxation time approximation to take into account the so-called side-jump effect, but we show that the ensuing correction happens to vanish for Bjorken symmetry. After expressing the conserved currents in terms of chiral moments, we derive equations of motion for these moments from the Boltzmann equation. After a suitable truncation, these equations allow us to study the transition from the early-time collisionless regime to the hydrodynamic regime at late time, where the parity-violating chiral moments decay exponentially. The truncation that we use for the parity-violating moments is shown to be identical to Israel-Stewart's 14-moment approximation. Our final set of equations  can be used to calculate the energy-momentum tensor, vector-, and axial-vector currents with chiral degrees of freedom for possible applications in heavy-ion collisions.
\end{abstract}

\maketitle

\section{Introduction}

Polarization phenomena in relativistic heavy-ion collision have recently triggered a large interest. For massive spin-1/2 particles, measurements of the polarization of $\Lambda$-hyperons~\cite{STAR:2017ckg,Adam:2018ivw,ALICE:2019aid,Mohanty:2021vbt} confirm that the spin is aligned along the local equilibrium fluid vorticity~\cite{Liang:2004ph,Voloshin:2004ha,Betz:2007kg,Becattini:2007sr}. Furthermore, for massless particles the (axial) chiral vortical effect and the chiral magnetic effect (chiral separation effect) are well known equilibrium phenomena, where the (axial) vector current gains contributions from vorticity~\cite{Vilenkin:1979ui,Vilenkin:1980zv,Erdmenger:2008rm,Banerjee:2008th,Son:2009tf} or magnetic fields~\cite{Kharzeev:2007jp,Fukushima:2008xe}, respectively, see, e.g., Refs.~\cite{Hosur:2013kxa,Kharzeev:2015znc,Huang:2015oca} for reviews and Refs.~\cite{Hidaka:2018ekt,Yang:2018lew,Huang:2020kik,Palermo:2023cup} for recent related work. While local-equilibrium effects for polarization seem nowadays well understood, only little insight has been gained into off-equilibrium dynamics of relativistic fluids with polarization. It is the purpose of the present paper to present progress in this direction. 

A commonly used tool to describe systems beyond local equilibrium is kinetic theory, e.g., as derived for chiral particles in Refs.~\cite{Son:2012wh,Stephanov:2012ki,Chen:2012ca,Manuel:2013zaa,Chen:2015gta,Hidaka:2016yjf,Hidaka:2017auj,Ebihara:2017suq} and for massive particles with spin in Refs.~\cite{Weickgenannt:2019dks,Gao:2019znl,Hattori:2019ahi,Wang:2019moi,Weickgenannt:2020aaf,Liu:2020flb,Weickgenannt:2021cuo,Sheng:2021kfc,Wagner:2022amr,Wagner:2023cct}. However, solving kinetic theory can be highly challenging. Therefore, it is common to turn to hydrodynamics instead, see, e.g., Refs.~\cite{Shi:2020htn,Speranza:2021bxf,Abboud:2023hos} for chiral hydrodynamics and Refs.~\cite{Florkowski:2017ruc,Florkowski:2017dyn,Florkowski:2018fap,Montenegro:2018bcf,Hattori:2019lfp,Bhadury:2020puc,Singh:2020rht,Montenegro:2020paq,Gallegos:2021bzp,Fukushima:2020ucl,Li:2020eon,Wang:2021ngp,Hu:2021pwh,Hongo:2021ona,Daher:2022xon,Weickgenannt:2022zxs,Weickgenannt:2022jes,Gallegos:2022jow,Cao:2022aku,Weickgenannt:2022qvh,Biswas:2023qsw,Weickgenannt:2023btk} for spin hydrodynamics. Although hydrodynamics is originally based on a gradient expansion around local equilibrium, it has recently been realized that it can give a very good description of the quark-gluon plasma created in heavy-ion collisions even before the system is close to local equilibrium~\cite{Heller:2015dha}. In the context of heavy-ion collisions, this can, to a large extent, be attributed to the fact that, in some formulation of hydrodynamics, the resulting equations of motion can capture the trivial dynamics of the early time, rapidly expanding and mostly collisionless, regime. Thus, in this paper, we shall rely on the analysis presented in  Ref.~\cite{Blaizot:2021cdv} (see also Refs.~\cite{Kurkela:2019set,Giacalone:2019ldn,Almaalol:2020rnu,Ambrus:2021sjg,Chattopadhyay:2021ive,Jaiswal:2021uvv,Jaiswal:2022udf} for related work)  in order to obtain equations of motion for a system of particles with chiral degrees of freedom and broken parity. This will allow us to consider off-equilibrium contributions to the vector and axial-vector currents, the latter being directly related to polarization.

In this paper, we consider a boost-invariant setup, i.e. we  consider Bjorken expansion~\cite{PhysRevD.27.140}, a commonly used model to mimic the expansion of the quark-gluon plasma in a relativistic heavy-ion collision. It is characterized by boost invariance along the collision axis ($z$-axis) and translational invariance in the transverse plane ($x$-$y$-plane). In contrast to previous works, we allow for parity breaking, and take into account a possible imbalance between left- and right-handed massless particles. Physically, this corresponds to a situation with initial net vector- and axial-vector currents, which are uniform in the transverse plane and decay as the system evolves. As a consequence, we deal with a new set of chiral moments, corresponding to the vector- and the axial-vector currents, as well as off-diagonal components of the energy-momentum tensor. We derive equations of motion for these chiral moments from kinetic theory and study their behaviour both in the fast expanding free-streaming regime,  and the hydrodynamic regime dominated by collisions. In the former case, we find that for generic initial conditions, the system quickly reaches its late time behavior characterized by a simple power law decay. In the hydrodynamic regime the spatial parts of the vector- and axial-vector currents and the nondiagonal parts of the energy-momentum tensor decay exponentially due to collisions. We also find that the equations of motion for the zeroth components of the vector- and axial-vector currents decouple from all other moments, respectively, and show ideal behavior.

The equations of motion for the moments which are part of the conserved currents couple to higher moments. In order to close the system of moment equations, one requires a reasonable truncation.
Following Ref.~\cite{Blaizot:2021cdv}, we consider a simple truncation and show that the effect of  the neglected higher moments can be accounted for  by a redefinition of some transport coefficients in the truncated equations of motion.  The resulting truncated set of equations is then shown to be able to capture the dynamics of the system during its full evolution towards the hydrodynamic regime. With this, we obtain, as the main result of this work,  a closed set of equations of motion for chiral hydrodynamics with Bjorken symmetry, valid both at early and late time of the expansion.

This paper is organized as follows. In Section \ref{cktsec}, we derive a relaxation-time approximation for chiral kinetic theory, which takes into account the so-called side-jump effect and is consistent with the requirement of covariance. Then, in Section \ref{bjorksec}, we implement Bjorken symmetry, simplifying the previously obtained kinetic equation. The result serves as the starting point to derive equations of motion for chiral hydrodynamics in the following. In Section \ref{momsec}, we express the charge and axial-charge currents as well as the energy-momentum tensor in terms of chiral moments, and  obtain equations of motion for the latter. Section \ref{freesec} is dedicated to the exact solutions of the equations for the chiral moments in the collisionless limit, focusing on their late time power law behaviors. In Section \ref{anasec}, we consider the full set of equations for the relevant chiral moments and check the accuracy of a simple truncation over the entire evolution. Closed equations of motion for all the components of the conserved currents are given. Finally, in Section \ref{compsec}, we compare our results to those obtained from the frequently used 14-moment approximation as originally suggested by Israel and Stewart. Conclusions are provided in Section \ref{concsec}. Throughout this paper, we use the following notation and conventions, $a\cdot b \equiv a_\mu b^\mu$, $g_{\mu\nu}=\text{diag}(+,-,-,-)$, $\epsilon^{0123}=-\epsilon_{0123}=1$. We do not distinguish between upper and lower spatial indices of three vectors.

\section{Covariant chiral kinetic theory and relaxation time approximation}
\label{cktsec}

In this work, we use the chiral kinetic theory, introduced in Ref.~\cite{Chen:2015gta} as the microscopic theory that serves as a starting point for the  derivation of  equations of motion for chiral hydrodynamics.
The equation of motion in kinetic theory is usually given by the Boltzmann equation, of the generic form
\begin{equation}
    p \cdot \partial f(x,p) = C[f]\; , \label{boltz}
\end{equation}
where $f(x,p)$ is the distribution function and $C[f]$ is the collision term. When  chiral particles are involved, subtleties arise, which we briefly discuss in this section, 

In order to take into account a possible imbalance between left- and right-handed particles, we define two distinct distribution functions $f^\lambda(x,p)$ with $\lambda=\pm1$ for the two different chiralities of massless spin-1/2 particles. It has been shown in Refs.~\cite{Chen:2014cla,Chen:2015gta} that these distribution functions do not transform as scalars under Lorentz transformations and  therefore depend  on the frame in which they are defined. This feature is related to the fact that, in the massless case, it is not possible to split the total angular momentum into spin and orbital parts in a frame-independent way~\cite{Chen:2015gta,Stone:2014fja,Stone:2015kla,Speranza:2020ilk}. In the following, we denote the four-velocity of the reference frame by the four-vector $n^\mu$, and the frame dependence of the respective distribution function by a subscript $n$. It is important to note that the distribution functions themselves are not measurable (beyond the classical approximation), and are therefore allowed to depend on a reference frame. On the other hand, we require any observable quantity to be frame independent.

 The frame dependence of the definition of spin angular momentum for massless particles results in the so-called side-jump effect~\cite{Chen:2015gta}, i.e., the form of the collision term also depends on the reference frame. In particular, a local collision (the particles meet at one single space-time point) in one reference frame, may appear nonlocal in a different frame. The form of the collision term in principle should be derived from the underlying quantum theory and will in general feature a complicated structure, c.f.\ Refs.~\cite{Weickgenannt:2020aaf,Weickgenannt:2021cuo,Sheng:2021kfc,Wagner:2022amr}. However, much insight can be gained by using a simplified collision term, employing the so-called relaxation time approximation (RTA). In the RTA, the collision term is proportional to the difference between the distribution function and its local-equilibrium value, divided by a relaxation time $\tau_R$. When using the RTA for chiral kinetic theory, one has to take into account the side-jump effect in the collision term and preserve covariance of observables. In the following, we will show how to do this. We shall work in a semi-classical approximation and expand all the relevant quantities $A$ (e.g. distribution functions, currents) up to first order in $\hbar$, viz.
\begin{equation}
    A=A^{(0)}+\hbar A^{(1)}+\mathcal{O}(\hbar^2)\; .
\end{equation}

We start by defining the left- and right-handed currents~\cite{Chen:2015gta}
\begin{equation}
    J_\lambda^\mu\equiv \int dP\, j_\lambda^\mu \label{currents}
\end{equation}
with $dP\equiv (d^3p/E_p)$ and, to first order in $\hbar$, $j_\lambda^\mu$ has the following structure
\begin{equation}
    j_\lambda^\mu=\left[p^\mu+\hbar S_n^{\mu\nu}(\partial_\nu+\mathcal{C}^\lambda_\nu)\right]f_n^\lambda \label{smallj}\; .
\end{equation}
Here, the dipole moment tensor for massless particles is defined as
\begin{equation}
  S^{\mu\nu}_n\equiv \frac{\lambda}{2p\cdot n}\epsilon^{\mu\nu\alpha\beta}p_\alpha n_\beta \; , \label{dipmom}
\end{equation}
where $n^\mu$ is the frame vector introduced above. Furthermore, $\mathcal{C}^\lambda_\nu$ in \eq\eqref{smallj} is the contribution of the collisions to the currents~\cite{Chen:2015gta,Hidaka:2016yjf}, called jump current in Ref~\cite{Chen:2015gta}. Given the structure of Eqs.~(\ref{smallj}) and (\ref{dipmom}), it is clear that the jump current vanishes if one chooses the frame vector to be $n^\mu\sim \mathcal{C}^\mu$. 

In order to estimate the collisional contributions to the chiral currents, we shall rely on the relaxation time approximation, implemented in such a way that the covariance (i.e. independence of $n^\mu$) of \eq\eqref{smallj} is guaranteed. In order to be consistent with the conservation of left- and right-handed particles, equivalent to the conservation of the vector- and axial-vector currents, we require the currents \eqref{currents} to be separately conserved, 
\begin{equation}
   \partial \cdot J_\lambda=0\; , \label{consva}
\end{equation}
while the divergences of the integrands are equal to a collision term, or equivalently,
\begin{equation}
    \partial\cdot j_\lambda = C_\lambda\; . \label{partialj}
\end{equation}
This equation is frame-independent. At zeroth order in the $\hbar$ expansion, \eq\eqref{partialj} reduces to
\begin{equation}
    p\cdot \partial f^{\lambda(0)}=C^{\lambda(0)}\; ,
\end{equation}
where we inserted \eq\eqref{smallj}. Therefore, $C_\lambda$ at zeroth order can be identified with the classical collision term in \eq\eqref{boltz}.
In order to ensure covariance of the currents \eqref{currents}, the distribution functions have to transform under Lorentz transformations as~\cite{Chen:2015gta}
\begin{equation}
    f_n^\lambda-f_{n^\prime}^\lambda= -\hbar S_n^{\mu\nu} \frac{n^\prime_\mu}{p\cdot n^\prime} (\partial_\nu+\mathcal{C}^\lambda_\nu) f^{\lambda(0)}\;. \label{ftransform}
\end{equation}
Indeed, in this case the difference $ \delta j_\lambda^\mu$  between the values of $j^\mu_\lambda$ in the frames characterized by $n^\mu$ and $n^{\prime\mu}$ becomes, up to first order in $\hbar$,
\begin{align}
    \delta j_\lambda^\mu &= -p^\mu \hbar S_n^{\rho\nu} \frac{n^\prime_\rho}{p\cdot n^\prime} (\partial_\nu+\mathcal{C}^\lambda_\nu) f^{\lambda(0)}+\hbar (S_n^{\mu\nu}-S_{n^\prime}^{\mu\nu})(\partial_\nu+\mathcal{C}^\lambda_\nu) f^{\lambda(0)}\n\\
    &= \hbar \epsilon^{\alpha\beta\rho\mu} \frac{p_\alpha n^\prime_\beta n_\rho}{2 (p\cdot n)(p\cdot n^\prime)} p\cdot (\partial+\mathcal{C}^\lambda)f^{\lambda(0)}\; ,
\end{align}
where the zeroth-order distribution function $f^{\lambda(0)}$ is independent of $n^\mu$.\footnote{In order to get the second line, we used the identity (Schouten identity) 
\beq
\epsilon^{\lambda\nu\alpha\beta}p^\mu+\epsilon^{\nu\alpha\beta\mu}p^\lambda+\epsilon^{\alpha\beta\mu\lambda}p^\nu+\epsilon^{\beta\mu\lambda\nu}p^\alpha+\epsilon^{\mu\lambda\nu\alpha}p^\beta=0.\eeq
}
Therefore, by requiring the frame independence of the currents, we find the following relation between $C_\lambda$ in \eq\eqref{partialj} and $\mathcal{C}^{\lambda(0)}$,
\begin{equation}
    p\cdot \partial f^{\lambda(0)}= C^{\lambda(0)}\equiv -p\cdot \mathcal{C}^{\lambda(0)}f^{\lambda(0)}\; . \label{boltz0}
\end{equation}
This also implies that if $f^{\lambda(0)}$ is not a solution of the Boltzmann equation, as it is the case in local equilibrium  ($p\cdot\partial f^{\rm eq}\ne 0$), the current is frame dependent. In this sense, the notion of local equilibrium is a frame dependent concept.

In order to find a simple expression for $C_\lambda$, it is natural to use the relaxation time approximation at zeroth order,
\begin{equation}
    p\cdot \partial f^{\lambda(0)}=p\cdot u \frac{f^{\lambda(0)}-f^{\lambda(0)}_\text{eq}}{\tau_R}\; , \label{rta0}
\end{equation}
where $f^{\lambda(0)}_\text{eq}$ is the zeroth-order local-equilibrium distribution function. Inserting this into \eq\eqref{boltz0}, we conclude that 
\begin{equation}
    \mathcal{C}_\nu^{\lambda(0)}= -u_\nu \frac{f^{\lambda(0)}-f^{\lambda(0)}_\text{eq}}{\tau_R f^{\lambda(0)}}\; .
\end{equation}
In principle, there could be also  a contribution parallel to $p_\nu$, however, this would vanish in \eq\eqref{smallj}, since $\mathcal{C}_\nu$ is contracted with $S_n^{\mu\nu}$. It is interesting to note that in the RTA, since $\mathcal{C}_\mu^\lambda$ is proportional to $u_\mu$, its contribution to the currents \eqref{currents} vanishes for $n^\mu=u^\mu$. This means that the jump current vanishes in the fluid rest frame, which therefore correspond to the``no-jump frame" of Ref.~\cite{Chen:2015gta}. A similar result has also been obtained in Ref.~\cite{Hidaka:2017auj} when considering a Boltzmann equation with a collision term involving 2-to-2 scattering.

We also need a reasonable approximation for the first order correction $C^{\lambda(1)}$. We require it to be  proportional to the difference between the first-order distribution function and its equilibrium value, and in addition to be frame-independent, since the left-hand side of \eq\eqref{partialj} also is. As a first attempt, we consider the following ansatz (see  also Ref.~\cite{Hidaka:2017auj})
\begin{equation}
    C^{\lambda(1)}=-p\cdot u \frac{f_n^{\lambda(1)}-f^{\lambda(1)}_{n,\text{eq}}}{\tau_R^\lambda}+ X_n \frac{f^{\lambda(0)}-f^{\lambda(0)}_\text{eq}}{\tau_R^\lambda}\; , \label{badansatz}
\end{equation}
where $X_n$ is an operator to be determined. By using  \eq\eqref{ftransform} for $f_n$ and $f_{n,\rm eq}$, one obtains the difference of the values of $C^{\lambda(1)}$ between two frames as
\begin{equation}\label{eqforXn}
    \delta C^{\lambda(1)}=\frac{p\cdot u}{\tau_R}   S_n^{\mu\nu} \frac{n^\prime_\mu}{p\cdot n^\prime} (\partial_\nu+\mathcal{C}^\lambda_\nu) \left(f^{\lambda(0)}-f^{\lambda(0)}_\text{eq}\right)+ (X_n-X_{n^\prime}) \frac{f^{\lambda(0)}-f^{\lambda(0)}_\text{eq}}{\tau_R}\; .
\end{equation}
Now taking $X_n\equiv -u_\mu   S_n^{\mu\nu} (\partial_\nu+\mathcal{C}_\nu^\lambda)$, we get \footnote{The motivation underlying this ansatz is as follows: In order to be able to cancel the first term in Eq.~(\ref{eqforXn}), $X_n$ has to be proportional to $\partial_\nu+\mathcal{C}^\lambda_\nu$. Furthermore, since it is related to the spin exchange in the frame $n$, it is natural to choose it proportional to $S^{\mu\nu}_n$. The only vector at our disposal to contract the open index is $u_\mu$, since $S_n^{\mu\nu}$ is orthogonal to both $p$ and $n$.}
\begin{equation}
X_n-X_{n^\prime}=-u_\mu\left(p^\mu  S_n^{\rho\nu} \frac{n^\prime_\rho}{p\cdot n^\prime}+\epsilon^{\alpha\beta\rho\mu} \frac{p_\alpha n^\prime_\beta n_\rho}{2 (p\cdot n)(p\cdot n^\prime)} p^\nu\right)(\partial_\nu+\mathcal{C}_\nu^\lambda)
\end{equation}
and therefore, assuming that $\tau_R$ is constant and $f^{\lambda(0)}$ is a solution of the Boltzmann equation,
\begin{equation}
    \delta C^{\lambda(1)}=  u_\mu \epsilon^{\alpha\beta\rho\mu} \frac{p_\alpha n^\prime_\beta n_\rho}{2 (p\cdot n)(p\cdot n^\prime)} p^\nu(\partial_\nu+\mathcal{C}_\nu^\lambda) \frac{f^{\lambda(0)}_\text{eq}}{\tau_R}=u_\mu \epsilon^{\alpha\beta\rho\mu} \frac{p_\alpha n^\prime_\beta n_\rho}{2 (p\cdot n)(p\cdot n^\prime)} p^\nu\left(\partial_\nu-u_\nu \frac{f^{\lambda(0)}-f^{\lambda(0)}_\text{eq}}{\tau_R f^{\lambda(0)}}\right) \frac{f^{\lambda(0)}_\text{eq}}{\tau_R}\; .
\end{equation}
Since this does not vanish, it means that the ansatz \eq\eqref{badansatz} leads to a frame-dependent form of $C_\lambda$, which does not fulfill our requirement. The difficulty is that local equilibrium breaks the frame independence of the currents, as we already saw from \eq\eqref{boltz0}. Thus local equilibrium can be defined only in a specific frame. In order to ensure frame independence of the currents near local equilibrium, we need to specify the frame choice in the relaxation time approximation and  force the distribution function to approach local equilibrium as defined in that frame. The only natural choice for this frame is the fluid rest frame, i.e., we define
\begin{equation}
    C^{\lambda(1)}=-p\cdot u \frac{f_n^{\lambda(1)}-f^{\lambda(1)}_{u,\text{eq}}}{\tau_R}+ X_n \frac{f^{\lambda(0)}}{\tau_R}\; , \label{rta1}
\end{equation}
where we also made use of $X_u=0$. Now we have
\begin{equation}
    \delta C_\lambda^{(1)}=\frac{p\cdot u}{\tau_R}   S_n^{\mu\nu} \frac{n^\prime_\mu}{p\cdot n^\prime} (\partial_\nu+\mathcal{C}^\lambda_\nu) f^{\lambda(0)}+ (X_n-X_{n^\prime}) \frac{f^{\lambda(0)}}{\tau_R}=0\, ,
\end{equation}
since $f$ is a solution of the Boltzmann equation. Thus, we obtained a covariant relaxation time approximation for chiral kinetic theory.

\section{Bjorken symmetry}
\label{bjorksec}

After having obtained an expression for the collision term which guarantees the frame independence of observables, we may specify $n^\mu$ without loss of generality. We will choose the fuid rest frame, i.e. $f\equiv f_u$ in the following.~\footnote{\bbb{By choosing the fluid rest frame we mean here that we define the distribution function in the frame which is at rest with the fluid.}} We aim at describing a system of chiral particles undergoing a longitudinal boost-invariant expansion in the $z$-direction, the so-called Bjorken flow~\cite{PhysRevD.27.140}. Furthermore, we assume translational invariance in the $x$-$y$-plane, such that $f^\lambda=f^\lambda(t,z,\p)$. However, we allow the momentum distribution to  break parity in the $x$-$y$-plane, i.e., $f(p_x)\neq f(-p_x)$, $f(p_y)\neq f(-p_y)$ in general. On the other hand, we assume parity invariance in $z$-direction, $f(p_z)= f(-p_z)$.

In order to implement boost invariance along the $z$-direction for the distribution functions $f_\lambda$, we require 
\begin{equation}
f^\lambda(t,z,\mathbf{p}_\bot,p_z)=f^{\prime\lambda}(\tau,\p_\bot,p_z^\prime)= f^{\lambda}(\tau,\p_\bot,p_z^\prime)+\hbar S_u^{\mu\nu} \frac{u^\prime_\mu}{p\cdot u^\prime} \partial_\nu f^{\lambda}(\tau,\p_\bot,p_z^\prime)
\end{equation}
with $\tau\equiv\sqrt{t^2-z^2}$, $p_z^\prime\equiv(p_z-E_pz/t)\gamma$,  $\gamma\equiv t/\tau$, $E_p\equiv p^0\equiv \sqrt{\p^2}$, $u^\mu\equiv(1/\tau)(t,0,0,z)$ and $u^{\prime\mu}\equiv(1,\boldsymbol{0})$. Here we \bbb{denoted the distribution functions defined in the frames corresponding to $u^\mu$ and $u^{\prime\mu}$ by $f$ and $f^\prime$, respectively, and} take into account the fact that $f^\lambda$ does not transform as a scalar under Lorentz transformations, but acquires an additional term from \eq\eqref{ftransform}. Using the explicit form of $S^{\mu\nu}$ in  \eq\eqref{dipmom}, we obtain
\begin{equation}
    f^\lambda(t,z,\mathbf{p}_\bot,p_z)= f^{\lambda}(\tau,\p_\bot,p_z^\prime)+\frac{\hbar}{2p^0(tp^0/\tau-z p_z/\tau)}\epsilon^{ij03}p^\prime_j \frac{z}{\tau} \partial_i f^{\lambda}(\tau,\p_\bot,p_z^\prime)=f^{\lambda}(\tau,\p_\bot,p_z^\prime)\; , \label{boostinv}
\end{equation}
where $f^\lambda$ is assumed to be independent of $x$ and $y$ \bbb{and hence the second term after the first equal sign vanishes. Thus, although the distribution function does not transform as a Lorentz scalar in general, it does here for the special case of a boost along the $z$-axis}.

One can verify that, for $n^\mu=u^\mu$, the term of $\mathcal{O}(\hbar)$ in \eq\eqref{smallj} does not contribute to the divergence of the current in  \eq\eqref{partialj}. To do so, we use the antisymmetry of the Levi-Civita tensor (hidden in $S^{\mu\nu}$), and note that neither $u^\mu$ nor $f$ depend on $x$ or $y$. Then using \eqs\eqref{rta0}, \eqref{rta1}, and \eqref{boostinv}, we find that
 the distribution functions for left- and right-handed particles follow the well known Boltzmann equation for Bjorken flow~\cite{Baym:1984np},~\footnote{\bbb{Note that due to the boost invariance we need to consider only the $z=0$-slice. All quantities are then automatically expressed in the fluid rest frame, which at the $z=0$-slice is identical to the lab frame.}}
\begin{equation}
    \left(\partial_\tau-\frac{p_z}{\tau}\partial_{p_z}\right) f^\lambda= -\frac{f^\lambda-f_\text{eq}^\lambda}{\tau_R}\; , \label{boltzbjor}
\end{equation}
where the local-equilibrium distribution function
\begin{equation}
    f^\lambda_\text{eq}=\frac{1}{2(2\pi)^3} e^{\beta(\mu_\lambda-\ E_p)}
\end{equation}
depends on the local effective temperature $T$ with $\beta\equiv 1/T$ and chemical potential $\mu_\lambda$, which are determined by matching conditions. Thus, the left- and right-handed chemical potentials are determined by the condition
\begin{equation}
    \int d^3p\, f^\lambda=\int d^3p\, f_\text{eq}^\lambda\; . \label{matchmu}
\end{equation}
Similarly, the temperature is determined by
\beq
\sum_{\lambda=\pm1}\int d^3p\, p^0 f^\lambda=\sum_{\lambda=\pm1}\int d^3p\, p^0 f_\text{eq}^\lambda\; .\label{temp}
\eeq
 In principle, the local-equilibrium distribution function may contain also a term proportional to $S_u^{\mu\nu}$ and the thermal vorticity~\cite{Chen:2015gta}.  However, the latter vanishes due to the assumption of translational invariance.

While it is possible to solve numerically Eq.~(\ref{boltzbjor}), in this paper we shall proceed by first deriving from it equations of motion for the physically relevant quantities (currents, energy-momentum tensor). These quantities are related to simple moments of the distribution function, and their equations of motion lend themselves to simple approximations which make them easy to solve, while providing a good representation of the exact solution.

\section{Moments and equations of motion}
\label{momsec}

In chiral hydrodynamics, we are interested in the vector ($J_V^\mu$) and axial-vector ($J_A^\mu$)  currents, 
\begin{equation}
    J_V^\mu\equiv \sum_{\lambda=\pm1} J_\lambda^\mu ,\qquad
    J_A^\mu\equiv \sum_{\lambda=\pm1} \lambda J_\lambda^\mu \; . \label{ja}
\end{equation}
As discussed above, to first order in $\hbar$, these currents are frame-independent, and we may therefore, without loss of generality, choose to express them in the fluid rest frame $n^\mu=u^\mu$. We get then
\begin{align}
J_V^\mu&= \sum_{\lambda=\pm1} \int dP\, \left(p^\mu+\hbar\lambda \epsilon^{\mu z\alpha0}\frac{p_\alpha}{2E_p}\partial_z\right)f^\lambda \;,\n\\
    J_A^\mu&= \sum_{\lambda=\pm1} \lambda\int dP\, \left(p^\mu+\hbar\lambda \epsilon^{\mu z\alpha0}\frac{p_\alpha}{2E_p}\partial_z\right)f^\lambda \; .
\end{align}
 Note that the temporal components $J_{A/V}^0$ have only equilibrium contributions, fixed by the matching conditions \eqref{matchmu}, while the spatial components are purely dissipative. 
In addition to the currents, we need to consider the canonical energy-momentum tensor. To first order in $\hbar$, this reads, in the fluid rest frame~\cite{Chen:2015gta}
\begin{equation}
    T^{\mu\nu}= \sum_{\lambda=\pm1} \int dP\, p^\nu \left(p^\mu+\hbar\lambda \epsilon^{\mu z\alpha0}\frac{p_\alpha}{2E_p}\partial_z\right)f^\lambda\; . \label{t}
\end{equation}

As mentioned earlier, rather than solving the kinetic equation for $f$, we shall transform it into a set of equations for moments of $f$. To this aim, we start by expressing the currents in terms of orthogonal chiral moments, given by integrals of spherical harmonics $Y_{n}^{\ell}(\theta,\phi)$.
Using $\partial_z f^\lambda=-(E_p/\tau)\partial_{p_z}f^\lambda$ and assuming that $f^\lambda$ is symmetric under $p_z\rightarrow -p_z$ and under the exchange of $p_x$ and $p_y$, we obtain the following expressions for the moments of the currents,
\begin{align}
    J_A^0&= \sum_{\lambda=\pm1}\lambda \mJ^\lambda_{00}\; ,& J_A^z&=0\; , & J_A^x&= -\sum_{\lambda=\pm1} \lambda \text{Re}   \mJ^\lambda_{11}= J_A^y\; , \n\\ 
    J_V^0&= \sum_{\lambda=\pm1}\mJ^\lambda_{00}\; ,& J_V^z&=0\; , & J_V^x&= -\sum_{\lambda=\pm1}\text{Re} \mJ^\lambda_{11}= J_V^y\; , \label{jmom}
\end{align}
and for those of the energy-momentum tensor
\begin{align}
T^{00}&=\sum_{\lambda=\pm1}\mT^\lambda_{00}\, , &
  T^{zz}  &= \sum_{\lambda=\pm1}  \left( \frac23 \mT^\lambda_{20}+\frac13\mT^\lambda_{00}\right)\;, \n\\ T^{xx}&=\frac13\sum_{\lambda=\pm1}\left(\mT^\lambda_{00}-\mT^\lambda_{20}\right)= T^{yy}\; , &  T^{xy}&= \frac16 \sum_{\lambda=\pm1}   \text{Im}\, \mT^\lambda_{22}= T^{yx}\; ,\n\\
 T^{xz}&= -\frac{1}{15}\frac\hbar\tau\sum_{\lambda=\pm1} \lambda\, \text{Re}  \left(4\mJ^\lambda_{11}+\mJ^\lambda_{31}\right)=- T^{yz}\; , &
 T^{zx}&= 0= T^{zy}\; . \label{tmom}
\end{align}
Note that $T^{i0}=T^{0i}=0$. 
Here, we have defined the chiral moments
\begin{align}
\mJ_{n\ell}^\lambda \equiv \int d^3p\, Y_n^\ell(\theta,\phi)  f^\lambda,\qquad 
    \mT_{n\ell}^\lambda \equiv \int d^3p\, p \, Y_n^\ell(\theta,\phi)  f^\lambda\; ,
    \label{momdefs}
\end{align}
where $Y_n^\ell(\theta,\phi)\equiv \Pc_n^\ell(\cos\theta) e^{i\ell\phi}$ is a spherical harmonic, $\Pc_n^\ell$ is an associated Legendre polynomial, $\theta$ is the polar angle with
$\cos\theta\equiv {p_z}/{p}$, $p\equiv E_p$
and $\phi$ is the azimuthal angle. 
The moments $\mT_{(2n)0}^\lambda$ are identical to the moments $\mathcal{L}_n$,  studied in previous works~\cite{Blaizot:2021cdv}. Additional  moments appear in \eqs\eqref{jmom} and \eqref{tmom}, which have not been considered before. They arise from the odd parity components of the momentum distribution and  chiral effects of order $\hbar$. Note in particular that parity breaking induces a dependence of the distribution function on the azimuthal angle $\phi$, hence the use of spherical harmonics instead of Legendre polynomials in the definition of moments. In the following, we derive the equations of motion for these new moments. Note that the equations of motion for the moments $\mT_{00}^\lambda$ and $\mT_{20}^\lambda$ are identical to those of the moments $\mathcal{L}_0$ and 
 $\mathcal{L}_1$ of Ref.~\cite{Blaizot:2017lht}. They will be recalled later (see Eqs.~(\ref{eq:T00T20}) below). 
 
 Using \eq\eqref{boltzbjor} and the properties of the associated Legendre polynomials, we obtain the equations of motion for the moments $\mathcal{J}_{n\ell}^\lambda$:
\begin{align}
  \partial_\tau \mathcal{J}_{n\ell}^\lambda  =&-\frac{1}{\tau}\left( a_{n\ell} \J_{n\ell}+b_{n\ell} \J_{(n-2)\ell}+ c_{n\ell} \J_{(n+2)\ell} \right)-\frac{\J_{n\ell}-\J_{n\ell,\text{eq}}}{\tau_R}\; , \label{dzdtau}
\end{align}
where the coefficients $a_{n\ell}$, $b_{n\ell}$, and $c_{n\ell}$ are given in Appendix \ref{coeffapp}.
They fulfill the relation
\begin{equation}
    {a}_{n\ell }+{b}_{n\ell}+{c}_{n\ell }=1+\ell \; . \label{abc1}
\end{equation}
We see that the equations of motion for different values of $\ell$, as well as those for even and odd powers of $n$ decouple from each other in the collisionless limit $\tau_R\rightarrow\infty$. Furthermore, we have $b_{00}=c_{00}=0$, so that the equation of motion for $\J_{00}$ decouples from all others. For $n+\ell$ even, we have the additional relation
\begin{equation}
     {a}_{n\ell }  \mathcal{P}^\ell_{n}(0)+{b}_{n\ell }\mathcal{P}^\ell_{n-2}(0)+{c}_{n\ell }\mathcal{P}^\ell_{n+2}(0)=\mathcal{P}^\ell_{n}(0)\; , \qquad \qquad n+\ell \text{ even}\; .
    \label{fun1}
\end{equation}
On the other hand, for $n+\ell$ odd we find
\begin{equation}
    {a}_{n\ell } (n+\ell) \mathcal{P}^\ell_{n-1}(0)+{b}_{n\ell }(n-2+\ell)\mathcal{P}^\ell_{n-3}(0)+{c}_{n\ell }(n+2+\ell)\mathcal{P}^\ell_{n+1}(0)=2(n+\ell)\mathcal{P}^\ell_{n-1}(0)\; , \qquad \qquad n+\ell \text{ odd}\; . \label{fun2}
\end{equation}

Similarly, the equation of motion for the moments $\mathcal{T}_{n\ell}^\lambda$ reads
\begin{align}
    \partial_\tau \mathcal{T}_{n\ell}^\lambda =&-\frac{1}{\tau}\left( \bar{a}_{n\ell} \T_{n\ell}+\bar{b}_{n\ell} \T_{(n-2)\ell}+ \bar{c}_{n\ell} \T_{(n+2)\ell} \right)-\frac{\T_{n\ell}-\T_{n\ell,\text{eq}}}{\tau_R}\; , \label{dxdtau}
\end{align}
with the coefficients (see Appendix \ref{coeffapp}) fulfilling the relation
\begin{align}
  \bar{a}_{n\ell }+\bar{b}_{n\ell }+\bar{c}_{n\ell }&=2+\ell\; , \label{abc0}
\end{align}
as well as a relation analogous to \eqref{fun1} for $n+\ell$ even, and \eqref{fun2} for $n+\ell$ odd, respectively.

\section{Collisionless limit}
\label{freesec}
The strategy for solving (approximately) the kinetic equation is to transform it into a truncated set of equations for appropriate moments of the distribution function. In the spirit of Ref.~\cite{Blaizot:2021cdv}, we look for a reduced set of coupled equations for the moments which are capable of describing both the early time collisionless regime and the late time, hydrodynamic regime. To that aim, in this section, we examine general features of the solutions of the moment equations in the collisionless regime. More technical details  are given in  Appendix \ref{freeapp}. Then we discuss a simple truncation that provides a reasonable approximation for the lowest two moments $\mathcal{J}_{11}^\lambda $ and $\mathcal{J}_{31}^\lambda $. 

In the collisionless limit,  all the chiral moments exhibit simple power law behaviors at early and late times, as discussed in \ Appendix \ref{freeapp}\footnote{Folowing the terminology of Ref.~\cite{Blaizot:2021cdv} we shall refer to these behaviors as fixed point behaviors, although the notion of fixed point appears only naturally in the context of the non linear equation that drives the diagonal components of the energy-momentum tensor. Thus, the late time power law of the moments will be referred to as the stable free streaming fixed point.} . By early and late time we mean respectively $\tau\ll \tau_0$ and $\tau\gg\tau_0$, where $\tau_0$ is the unique time scale that appears in the collisionless regime ($\tau_0$ will be chosen as the initial time of the evolution, i.e., the time at which one fixes the initial values of the moments). These simple regimes, which are deduced in Appendix~\ref{freeapp} from the known form of the free streaming solution of the kinetic equation, emerge in the solution of the coupled equations for the moments thanks to the relations \eqref{abc1} and \eqref{abc0} on the one hand, and the relations \eqref{fun1} and \eqref{fun2} on the other hand. Note however that these regimes may be only approximately reproduced when a truncation of the moment equations is made. 

Consider the early time behavior, $\tau\ll \tau_0$. For $\ell=0$,  all the moments with $n$ odd are equal, and so are the even ones, which follows from Eq.~(\ref{abc1}). Furthermore, Eq.~(\ref{dzdtau}) indicates  that all these moments diverge as $\tau_0/\tau$ when $\tau\to 0$. Similarly,  from Eqs.~(\ref{abc0}) and (\ref{dxdtau}) we deduce that  the $\T_{n\ell}$ diverge as $(\tau_0/\tau)^2$.  For $\ell>0$, one needs to take into account that $\Pc_n^\ell(\cos\theta)=0$ for $n<\ell$. Thus, for $n=\ell$ or $n=\ell+1$, the terms $\propto b_{n\ell}$ in Eq.~(\ref{dzdtau})  or $\propto \bar{b}_{n\ell}$ in Eq.~(\ref{dxdtau}) do not contribute to the equations of motion. It follows that the moments with $\ell>0$ are proportional to each other,  $\J_{n\ell}/\J_{m\ell}=B_{n\ell}/B_{m\ell}$, with the proportionality constants  $B_{n\ell}$ satisfying the following relations (cf.\ Appendix \ref{freeapp})
\begin{align}
    B_{n\ell} a_{n\ell}+B_{(n-2)\ell} b_{n\ell}+B_{(n+2)\ell}c_{n\ell}&= 1-\ell\; ,\n\\
   B_{n\ell} \bar{a}_{n\ell}+B_{(n-2)\ell} \bar{b}_{n\ell}+B_{(n+2)\ell}\bar{c}_{n\ell}&= 2-\ell \; , \label{anotherrelation}
\end{align}
where 
\begin{equation}
    B_{n\ell}\equiv \lim_{x\rightarrow 1} \frac{\Pc_n^\ell(x)}{(1-x^2)^{\ell/2}}
    \label{bnell}
\end{equation}
is a finite number.
Note that, for $\ell=0$, the two equations (\ref{anotherrelation})  are identical to \eqs\eqref{abc1} and \eqref{abc0}, respectively.  

Similar considerations apply to the late time regime, $\tau\gg \tau_0$, whose properties are governed by  \eqs\eqref{fun1} and \eqref{fun2}. There we have $\J_{n\ell}/\J_{m\ell}=\Pc_n^\ell(0)/\Pc_m^\ell(0)$ for $n+\ell$ and $m+\ell$ even, and $\J_{n\ell}/\J_{m\ell}=[(n+\ell)\Pc_{n-1}^\ell(0)/(m+\ell)\Pc_{m-1}^\ell(0)]$ for $n+\ell$ and $m+\ell$ odd. All chiral moments 
with $n+\ell$ even decay as $\tau^{-1}$, while those with $n+\ell$ odd decay as $\tau^{-2}$, as can be deduced from \ \eqs\eqref{fun1} and \eqref{fun2}. The  moments $\T_{n\ell}$ exhibit a similar behavior (see Appendix \ref{freeapp}).

We focus now on the equations of motion for the moments $\J_{11}$ and $\J_{31}$, which are involved in the calculation of the currents and  of some of the components of the energy momentum tensor (see \eqs\eqref{jmom} and \eqref{tmom}).  These equations are the first in an infinite set of coupled equations. Our goal is to find a suitable truncation of this infinite set of equations, which preserves the essential features of the collisionless dynamics, in particular its late time behavior. The equations of motion for the other moments in \eqs\eqref{jmom} and \eqref{tmom} can be treated along similar lines, as will be discussed in the next section. 

The equation for $\mathcal{J}_{11}^\lambda$ reads
\begin{align}
  \partial_\tau \mathcal{J}_{11}^\lambda  =&-\frac{1}{\tau}\left( a_{11} \J_{11}+ c_{11} \J_{31} \right)\; , \label{dzdtau2}
\end{align}
with $a_{11}=4/5$ and $c_{11}=-2/15$.
This equation  couples  $\mathcal{J}_{11}^\lambda$ to $\mathcal{J}_{31}^\lambda$. The next equation in the hierarchy, that for $\mathcal{J}_{31}^\lambda$, couples $\mathcal{J}_{31}^\lambda$ to $\mathcal{J}_{11}^\lambda$ and $\mathcal{J}_{51}^\lambda$, and so on. 
More generally, we can write the set of equations of motion up to that for  $\J_{(2n-1)1}$, with $n\ge 1$, in the following matrix form, 
\begin{align}
\tau\partial_\tau\Vec{\mJ}^\lambda&=\mathcal{H} \Vec{\mJ}^\lambda\; 
\end{align}
with $(\Vec{\mJ}^\lambda)_q\equiv\J_{(2q-1)1}$,  $\mathcal{H}_{qq'}\equiv -\delta_{qq'} a_{(2q-1)1}-\delta_{q(q'+1)}b_{(2q-1)1}-\delta_{q(q'-1)}c_{(2q-1)1}$ and $1\le q\le n$.  The exact solution is reproduced by keeping all moments, that is, in the limit $n\to \infty$, where $n$ is the dimension of the matrix $\mathcal{H}$. In practice a truncation at a finite value of $n$ is necessary. The simplest of these truncations, which we refer to as the ``naive truncation'' consists in the following. For a finite value of $n$, the last equation, that for  $\J_{(2n-1)1}$, involves the moment  $\J_{(2n+1)1}$. The naive truncation  consists in closing the hierarchy by setting equal to zero this moment $\J_{(2n+1)1}$. Then we are left with a finite dimensional $n\times n$ linear problem which is easily solved.  

The eigenvalues of $\mathcal{H}$  depend on $n$, and in particular on the parity of $n$. Indeed, it turns out that odd truncations, that is truncations with $n$ odd, have one real eigenvalue. This eigenvalue is equal to -1 in the limit $n\to\infty$.  The components  of the corresponding eigenvector satisfy the relation  $\J_{(2q-1)1}=\mathcal{P}_{(2q-1)}(0)\J_{11}$ with $q\le n$,  which follows from Eqs.~\eqref{fun2} and 
\eqref{xlate}. They decay as $\tau^{-1}$ at large $\tau$, as expected of the exact behavior in the collisionless regime.  On the other hand, truncations with $n$ even have only pairs of complex conjugate eigenvalues, and do not allow us to correctly reproduce the late time behavior of $\J_{11}$. For this reason we limit ourselves to odd truncations in the following. 

In the case of $\J_{11}$ the lowest order naive truncation ($n=1$)  consists in dropping $\J_{31}$ in Eq.(\ref{dzdtau2}).  However this does not preserve the correct late time behavior since $a_{11}\ne 1$.  One could be tempted to solve the coupled equations for both moments $\J_{11}$ and $\J_{31}$, dropping $\J_{51}$ in the equation for $\J_{31}$. But this corresponds to an even ($n=2$) truncation which also 
 spoils the late time behavior, as we have argued. The next possible truncation is $n=3$, in which $\J_{71}$ is set to zero. One may expect that by increasing $n$ the naive truncation will provide a reasonably accurate representation of the exact solution, at least for the lowest moments. However the convergence appears to be slow, and non uniform. We shall therefore adopt a different strategy. 
\begin{figure}
    \centering
    \includegraphics[width=0.7\textwidth]{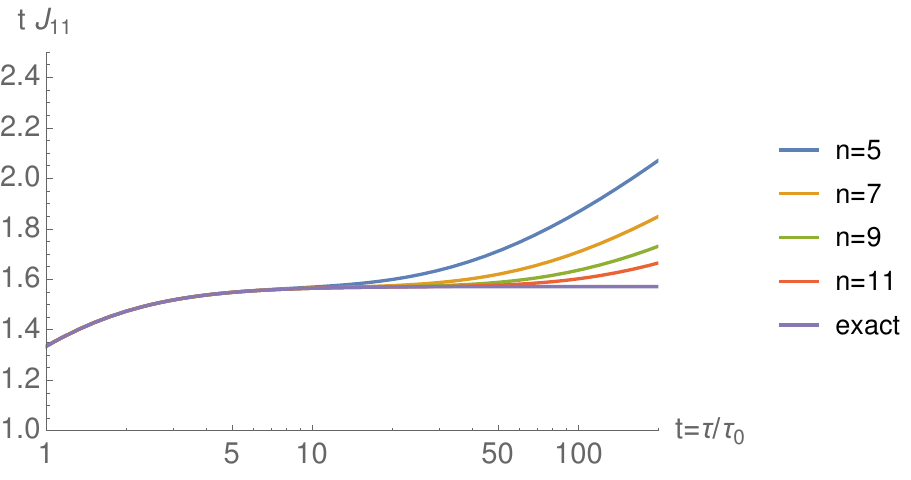}
    \caption{The quantity  $t\times \J_{11}(w)$, with $t\equiv \tau/\tau_0$,  obtained from the exact solution, and compared to the solution of a finite set of $n$ moment equations.  }
    \label{fig1}
\end{figure}
 
  In fact, what we need is a reasonable approximation for the late time behavior of the moments. That is, we need an approximation valid at times $\tau\gg \tau_0$, in particular in the case where $\tau_R\gg \tau_0$  so that the collisionless regime is fully developed before collisions start to act (if $\tau_0$ is close to $\tau_R$, the system enters quickly the hydro regime and the collisionless regime does not play much of a role). In that perspective one may view the effect of the higher moments as simply correcting the value of the moment $\J_{31}$ that appears in  the equation for $\J_{11}$, in such a way as to preserve its collisionless fixed point behavior for $\J_{11}$. Thus we are led to the following ansatz for $\J_{31}$
 \begin{align}
    \mathcal{J}_{31}^\lambda \simeq&-\frac32\J_{11}\; . \label{tau2const00}
\end{align}
 This leads to the following equation for $\J_{11}$:
 \begin{align}
   \partial_\tau \mJ_{11}^\lambda =& - \bbb{\frac{1}{\tau}}\J_{11}, \label{tau2const0}
\end{align}
whose solution is simply the fixed point behavior $\J_{11}\sim \tau^{-1}$.
\begin{figure}[h]
    \centering
    \includegraphics[width=0.7\textwidth]{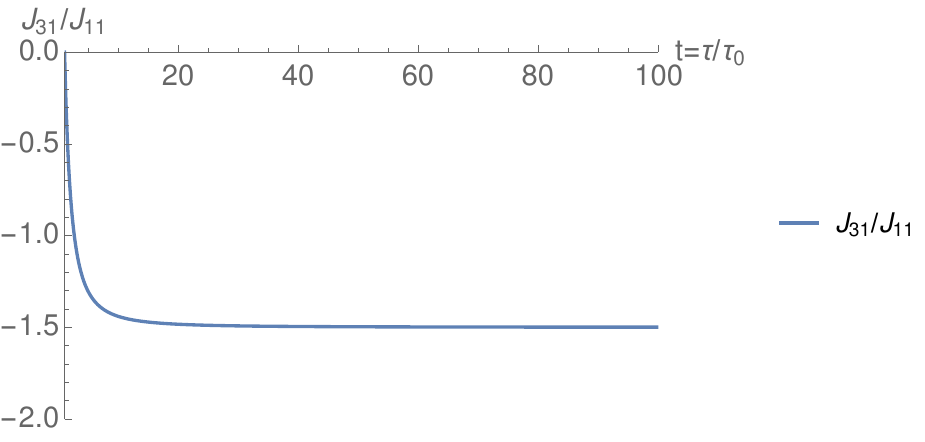}
    \caption{The ratio  $\J_{31}(t)/\J_{11}(t)$ obtained from the exact solution, as a function of $t=\tau/\tau_0$.}
    \label{figJ11}
\end{figure}
Thus, by forcing the relation between $\J_{31}$ and $\J_{11}$ to be that valid in the vicinity of the collisionless fixed point, we recover trivially the fixed point behavior of $\J_{11}$. As already mentioned, this procedure amounts to a simple renormalization of the coefficient in front of $\J_{11}$ in Eq.~(\ref{dzdtau2}): $a_{11}=4/5\mapsto 4/5+ (-3/2)\times (-2/15)=1$, where we have used $c_{11}=-2/15$. \bbb{Note that the purpose of this  renormalization  is to compensate for the error induced by the truncation of the moment equations at finite order. If  instead we were solving  the full set of moment equations, a renormalization would not be needed.}
To check that \bbb{\eq\eqref{tau2const0}} is a reasonable approximation, we compare the solution of Eq.~(\ref{tau2const0}) with the exact solution obtained by using the initial condition described in Appendix~\ref{freeapp}. 
\bbb{In Fig.~\ref{fig1}, we show the exact solution for $t\J_{11}$ and the solutions of different finite sets of moment equations.}
One sees that, for the chosen initial condition,  the collisionless fixed point is quickly reached \bbb{by the exact solution}, already for time $\tau\gtrsim 3\tau_0$ (recall that at the collisionless fixed point $t\J_{11}$  is independent of time). The behavior of the solution is well reproduced with a finite set of moments, at least up to some time (for $\tau\lesssim 100 \,\tau_0$ a good convergence is reached with $n=11$ moments. For larger times, a comparable convergence would require a higher number of moments.). Of course, the solution of Eq.~(\ref{tau2const0}) ignores the transient regime between the initial time $\tau_0$ and the time at which the $1/\tau$ behavior is reached: if it would be plotted in Fig.~\ref{fig1}, it would be an horizontal line starting at the initial value. But the early transient regime is not important for the present discussion. The same remark applies to the  ratio $\J_{31}/\J_{11}$ which is displayed in Fig.~\ref{figJ11}.  One sees that, after a relatively short transient regime, this ratio converges towards its fixed point value given in Eq.~(\ref{tau2const00}). In fact, we can use this fixed point value as an approximation for 
  $\J_{31}$ itself, not only in the right-hand side of Eq.~(\ref{tau2const0}). This is because  $\J_{31}$ enters the energy-momentum tensor as a first order correction in $\hbar$, so that a small error on  $\J_{31}$ will have little consequence anyway.

\section{Analysis of equations of motion and truncation}
\label{anasec}

We now extend our analysis to all the moments that appear in the expressions \eqref{jmom} and \eqref{tmom}, and include  the effect of collisions.  Closed equations of motion will be obtained by using a strategy similar to that used in the previous section to truncate the hierarchy.  This closed set of equations constitutes an effective theory for just the relevant moments, which include in particular all the hydrodynamic fields. This effective theory describes the transition from the early time collisionless regime, to the hydrodynamic regime at late time. Note that there are two time scales, the initial time $\tau_0$ and the collision time $\tau_R$. We shall assume in the present discussion that $\tau_0\ll\tau_R$, so that there is enough time for a collisionless regime to fully develop.  

We start with the equations of motion for $\J_{00}$ which is involved in the calculation of the axial and vector charges (see Eqs.~(\ref{jmom})):
\begin{align}
  \partial_w \J_{00}&=-\frac{1}{w}\J_{00}\; ,
     \label{zyreoms1}
\end{align}
where we have set $w\equiv\tau/\tau_R$. Furthermore, we used the relation $\J_{00}=\J_{00,\text{eq}}$, which follows from the matching condition \eqref{matchmu}. Since  this equation is closed already, no truncation is needed in this case, and the solution is simply $\J_{00}(w)\sim 1/w$.

Let us turn now to the equation for $\T_{22}$:
\begin{align}
    \partial_w \T_{22} =&-\frac{1}{w}\left( \frac67\T_{22}-\frac{2}{35}  \T_{42} \right)-\T_{22}\; ,
     \label{zyreoms2}
\end{align}
where we  used the fact that $\T_{22}$ vanishes in equilibrium.
At early time, $w\ll 1$, the last term in the right-hand side, which represents the effect of the collisions, is much smaller than the first one, which describes the expansion. Since we know that the initial collisionless dynamics quickly bring the moments to their free streaming stable fixed point, following Ref.~\cite{Blaizot:2021cdv}, and also along the same lines as in the previous section,  we approximate $\T_{42}$ in the right-hand side of Eq.~(\ref{zyreoms2}) by its value at the stable free streaming fixed point, i.e., we set
\begin{align}
\T_{42}&\simeq \frac{\mathcal{P}^2_{4}(0)}{\Pc^2_2(0)}\T_{22}\;  . \label{fptrun}
\end{align}
This guarantees that, in the absence of collisions, the moment $\T_{22}$ will have the correct  behavior for $\tau\gg \tau_0$. Again, this can be viewed as a renormalization of the coefficient of $\T_{22}$ in Eq.~(\ref{zyreoms2}): $6/7 \mapsto 1$.
With this identification, the equation for $\T_{22}$ in the absence of collisions becomes indeed 
  $\partial_w \T_{22}\simeq-\frac1w \T_{22}$,
whose solution is $\T_{22}\sim 1/w$, as expected near the stable free streaming fixed point (see  Section \ref{freesec}).

At late times ($w\gg 1$) the collision term dominates the evolution of the system and the equation for $\T_{22}$ becomes
\begin{align}
  \partial_w \T_{22} &\simeq-\T_{22}\; .
\end{align}
with the solution $\T_{22}\sim e^{-w}$.
Thus, the dissipative moments decay exponentially when approaching equilibrium. The reason for this exponential decay is the absence of equilibrium contributions in the equations of motion. This also  implies that there is no expansion in inverse powers of $w$ (gradient expansion) for these moments around local equilibrium.

The complete equation that accounts for the transition region  between the vicinity of the free-streaming fixed point and  the hydrodynamic regime, is therefore simply
\begin{align}
  \partial_w \T_{22}&\approx-\frac1w\T_{22}-\T_{22}\; .\label{zxkintot}
\end{align}
The solution is
\begin{align}
    \T_{22}&\sim w^{-1}e^{-w}\;   .
\end{align}
This solution exhibits nicely the transition from the collisionless regime where the first factor $\sim 1/w$ dominates,  to the collision dominated regime with exponential damping.

The equation for the moment  $\J_{11}$ was already discussed in the previous section. With  collisions, this equation becomes 
\begin{align}
   \partial_w \mJ_{11}^\lambda =&-\frac{1}{w}\J_{11} -\J_{11}\; , \label{tau2const}
\end{align}
an  equation which has the same structure as the equation for $ \T_{22}$, Eq.~(\ref{zxkintot}), that we have just discussed. Its solution reads
\begin{equation}
    \J_{11}\sim w^{-1}e^{-w}\; .
\end{equation}
\begin{figure}
    \centering
    \includegraphics[width=0.75\textwidth]{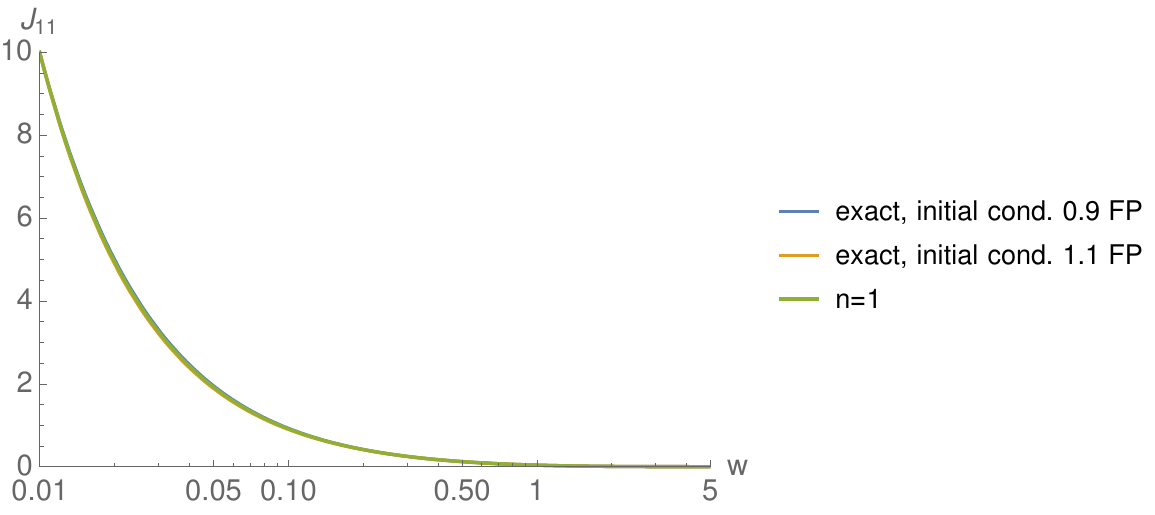}
    \caption{Comparison of $\J_{11}(w)$, as obtained from the lowest-order fixed-point truncation, Eq.~(\ref{tau2const}), to the exact solution obtained here by solving the $n$-moment equations with $n=11$. The vertical scale is arbitrary, with the initial value of $\J_{11}$ chosen to be 10. The initial values of the other moments are taken to be either their collisionless fixed point values, or to differ from those by $\pm 10\% $. As can be seen, the agreement with the exact solution is almost perfect and the sensitivity to the initial value of the moments is  very weak.}
    \label{fig5}
\end{figure}

 \begin{figure}
         \centering
    \includegraphics[width=\textwidth]{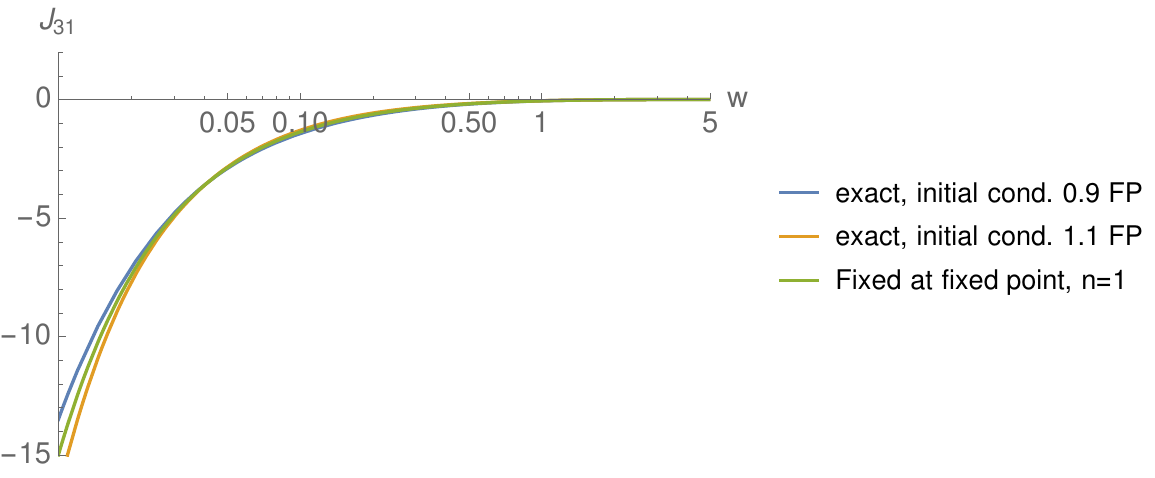}
        \caption{Comparison between the exact solution (obtained with $n=11$ moments) and the approximate solution $\J_{31}=(-3/2)\J_{11} $ (``fixed at fixed point", corresponding to what is referred to as ``attractor solution'' in \cite{Blaizot:2021cdv}): the agreement is perfect. Solutions with initial conditions where the moments differ from their fixed point values by $\pm 10\%$ approach the attractor very quickly.}
        \label{fig6}
\end{figure}
In Figure \ref{fig5} we plot the solution for $\J_{11}$ obtained from the fixed-point truncation at $n=1$ compared to the exact solution (i.e., the solution for $n=11$ \footnote{Note that the exponential damping makes the  convergence of the moment expansion considerably better than in the collisionless regime. \bbb{Thus, we may, in contrast to the free-streaming case, consider the solution for $n=11$ as the exact solution instead of solving the Boltzmann equation itself.}}), with different initial conditions. We see that the dependence on the initial conditions is very weak and already the lowest-order truncation is almost identical to the exact solution. Furthermore, we show in Figure \ref{fig6} the moment $\J_{31}$ obtained from $\J_{31}=-(3/2)\J_{11}$ (see Eq.~\eqref{tau2const0})  compared to the exact solution. If the system is initialized at the fixed point, the curves are identical. The sensitivity to the choice of initial condition is rather modest and quickly disappears as time increases. 

With the results obtained above and the equations of motion for $\T_{00}$ and $\T_{20}$ derived already in previous work~\cite{Blaizot:2021cdv}, we are ready to write down the  closed set of equations for chiral hydrodynamics which capture the transition from the collisionless fixed point to the hydrodynamic regime. Note first that the second term in \eq\eqref{t}, of order $\hbar$,  does not contribute to the energy-momentum conservation,
\begin{equation}
    \partial_\mu T^{\mu\nu}= \sum_{\lambda=\pm1} \int dP\, p^\nu p\cdot\partial f^\lambda\; ,
\end{equation}
since $\partial_x E_p=\partial_y E_p=0$ as a consequence of  the symmetry of the present setup. Therefore, we can ensure energy-momentum conservation by using the Landau matching conditions for the dominant part of the energy-momentum tensor
\begin{equation}
    u_\mu \sum_{\lambda=\pm1} \int dP\, p^\nu p^\mu f^\lambda = u_\mu \sum_{\lambda=\pm1} \int dP\, p^\nu p^\mu f^\lambda_\text{eq}\; ,
\end{equation}
 identically to what was done without chiral degrees of freedom in Ref.~\cite{Blaizot:2021cdv}. Thus, the equations of motion for the diagonal components of the energy-momentum tensor can be taken from these references without modifications~\cite{Blaizot:2021cdv},\footnote{In the notation of Ref.~\cite{Blaizot:2021cdv} $\sum_\lambda\T_{00}\equiv \mathcal{L}_0$ and $\sum_\lambda\T_{20}\equiv \mathcal{L}_1$. }
 \begin{align}\label{eq:T00T20}
     \partial_w \T_{00}&=-\frac1w \left(\frac43\T_{00}+\frac23\T_{20} \right)\; ,\n\\
     \partial_w \T_{20}&=-\frac1w\left(\frac{31}{15}\T_{20}+\frac{8}{15}\T_{00}\right)-\T_{20}\; ,
 \end{align}
 where the factor $31/15$ in the second equation includes the same kind of renormalization as discussed above ($38/21\mapsto 31/15$, see \cite{Blaizot:2021cdv}). Note that since $\T_{00}$ is nonzero in local equilibrium, both $\T_{00}$ and $\T_{20}$ decay with power laws in the hydrodynamic regime. In contrast to $\J_{00}$, for $\T_{00}$ the exponent of this power law differs between the free-streaming regime with $\T_{00}\sim \tau^{-1}$ and the hydrodynamic regime with $\T_{00}\sim \tau^{-4/3}$.
 
 The equations of motion for the remaining components of the chiral currents are obtained by inserting \eqs\eqref{zxkintot} and \eqref{tau2const} into \eqs\eqref{jmom} and \eqref{tmom}. The set of equations of motion for chiral hydrodynamics then reads
\begin{align}
\partial_w J_A^0&=-\frac1w J^0_A\; , \n \\
    \partial_w J_A^x &= -\frac{1}{w} J_A^x-J_A^x\; ,  \n\\
     \partial_w T^{xx}&=-\frac1w  T^{xx}-T^{xx} \; , \n\\
     \partial_w T^{xz}&=-\frac2w T^{xz}-T^{xz} \n\\
    \partial_w T^{xy}&=-\frac1wT^{xy}-T^{xy}\; .
    \label{eomfin}
\end{align}
The equations of motion for $J_V^\mu$ are identical to those for $J_A^\mu$. The set of equations \eqref{eomfin} can be solved to obtain the (axial-) charge current and energy-momentum tensor for a chiral fluid at any time of the Bjorken expansion.
A comparison of the results obtained by solving these equations to the exact solution shows the same quality as that made earlier for $\J_{11}$ and $\J_{31}$ (see Figs.~\ref{fig5} and \ref{fig6}).

We conclude this section by summarizing the two main results that we have obtained so far. First, from Eq.~(\ref{zyreoms1}) we find that the (axial-) charge density, given by a linear combination of $\J_{00}$ with $\lambda=\pm 1$ (see Eqs.~(\ref{jmom})), decouples from all other moments and decays $1/ \tau$ during the full evolution, from free streaming throughout the transition to hydrodynamics. Second, the spatial components of the (axial-) vector current and the nondiagonal components of the energy-momentum tensor decay as power laws for free streaming, but exponentially in the hydrodynamic regime. This exponential decay results in a very quick damping of the relevant moments. Therefore, any initial values of these moments will have disappeared at freeze out in our setup. On the other hand, if vorticity or magnetic fields were included in the formalism, they would lead to equilibrium contributions to the chiral currents due to the chiral vortical and chiral magnetic effects~\cite{Vilenkin:1979ui,Vilenkin:1980zv,Erdmenger:2008rm,Banerjee:2008th,Son:2009tf,Kharzeev:2007jp,Fukushima:2008xe}, respectively. \bbb{The equilibrium currents then would appear linearly on the right-hand sides of each equation in \eqref{eomfin}, rendering them inhomogeneous differential equations and preventing the exponential decay at late time.} In that case, all currents would decay as power laws in the hydrodynamic regime, and hence survive much longer. In turn, any measurement, in heavy-ion collisions, of non vanishing spatial components of the (axial-) vector current or of nondiagonal components of the energy-momentum tensor, would be an indication of  the presence of either vorticity or electromagnetic fields, instead of being a mere consequence of specific initial condition. Further investigation of these effects beyond equilibrium is left for future work.

\section{Comparison to 14-moment approximation}
\label{compsec}

An interesting connection can be observed when comparing the results of Section \ref{anasec} to those of the Israel-Stewart 14-moment approximation~\cite{Israel:1979wp}. For the latter, only the 14 moments, those which contribute to the charge current and to energy-momentum tensor, are taken into account, while the moments of rank 3 or higher in the spatial momentum are simply neglected\footnote{Strictly speaking, one should call this approximation ``$2\times14$-moment approximation" in the chiral case, taking into account the  moments of both chiralities.}.  Here we merely point out the connection between our present approach and that based on the 14-moment approximation,  and refer to Appendix \ref{14app} for a more complete discussion. Consider for instance the moment $\T_{42}$. This is given by 
\begin{align}
    \T_{42}&= \frac{15}{2}\int d^3p\, E_p \left(7\cos^2\theta-1\right)\sin^2\theta\, e^{2i\phi} f^\lambda\simeq -\frac52 \T_{22}=\frac{\Pc^2_4(0)}{\Pc^2_2(0)}\T_{22}\; ,
\end{align}
which is the same as in the fixed-point truncation \eqref{fptrun}. Analogous relations are valid for all other moments, as long as the lowest-order truncation is used and the moments vanish in equilibrium.
 We conclude that in order to reproduce the correct free-streaming fixed points within the lowest-order truncation, one should simply drop all terms which depend on higher orders of $p_z/E_p$ in the Legendre polynomials rather than approximating them. The reason is that the term $p_z\tau/\tau_0$ in the distribution function \eqref{freesol} in the free-streaming regime suppresses all higher-order terms of the Legendre polynomials, $\Pc^\ell_n(p_z/E_p)\sim \Pc^\ell_n(0)$ for $n+\ell$ even, or $\Pc_n^\ell(p_z/E_p)\sim \Pc^\ell_{n-1}(0)$ for $n+\ell$ odd. Therefore, the Bjorken expansion leads to vanishing higher-order moments, see also Appendix \ref{14app}. This behavior is the same as for the usual moments without chirality. However, as soon as coupled equations of motion for more than one moment are considered, as, e.g., is the case for the moments $\mathcal{T}^\lambda_{00}$ and $\mathcal{T}^\lambda_{20}$, which both appear in the energy-momentum tensor, the situation is different. In this case, the truncation is not unique, and the nonorthogonality of the Israel-Stewart moment starts to play a role. Furthermore, for moments which are nonzero at the hydrodynamic fixed point, equilibrium contributions have to be taken care of. We refer to Appendix \ref{14app} for a more detailed discussion.

\section{Conclusions}
\label{concsec}

In this paper, we studied the equations of motion for chiral hydrodynamics derived from kinetic theory in a boost-invariant expanding system. We showed that in chiral kinetic theory the usual relaxation-time approximation has to be modified in order to take into account the side-jump effect. In this approach, we found that the no-jump frame, in which the contribution from the side-jump effect to the chiral currents vanishes, is given by the fluid rest frame. We also pointed out that the concept of local equilibrium is frame dependent, whereas observables are frame independent, as long as they are obtained from a distribution function which is a solution of the Boltzmann equation. For convenience, we chose the fluid rest frame for the derivation of the equations of motion. We then showed that imposing Bjorken symmetry significantly simplifies the Boltzmann equation, formally reducing it to its well-known form in the absence of chiral degrees of freedom. 

We derived equations of motion for the chiral moments, which correspond to linear combinations of the components of the conserved currents. These  moments can be written as integrals over the momentum angles expressed through spherical harmonics $Y_n^\ell(\theta,\phi)$, and weighted by the distribution function and  some fixed power of the momentum. The equations of motion couple only chiral moments with the same powers of $p$, the same $\ell$, and either even or odd values of $n$, respectively. The coefficients in each set of equations of motion fulfill two important relations, corresponding to the two free-streaming fixed points associated respectively to the early and late time behaviors of the moments in the collisionless regime. We analyzed these fixed points by studying the exact solutions for the chiral moments in the collisionless regime. In particular, near the stable fixed point, moments with $n+\ell$ even were shown to decay as $1/\tau$, while those with $n+\ell$ odd decay  faster by an additional power $1/\tau$. We also found that naive truncations of the equations, consisting in dropping moments beyond a certain order, converge only slowly, and yield results that depend on whether an even or odd number of moments is kept. 

We then discussed the full equations of motion. A special case is given by those for $\J_{00}$, corresponding to charge and axial-charge densities. These quantities decouple from all other chiral moments in the exact equations of motion. Taking into account the matching conditions, these equations of motion are identical to those of ideal chiral hydrodynamics at any time of the expansion, even in the free-streaming regime. In the other equations of motion, a truncation procedure is needed in order to close the respective set of equations. We argued that in the vicinity of the stable free-streaming fixed point, we can take into account the effects of the neglected higher moments by a redefinition of a transport coefficient, resulting in the decay of the chiral moments with the correct power law at the stable fixed point.  In the hydrodynamic regime all parity-violating moments decay exponentially, so that the precise value of the  coefficients of the free parts of the equations of motion become unimportant. Thus,  in the hydrodynamic regime,  the redefinition of transport coefficients does not affect the late-time behavior of the conserved currents. The modified equations of motion then provide a very good agreement with the exact solution throughout the evolution from the vicinity of the free-streaming fixed point to the hydrodynamic regime. Using these truncations, we obtained a closed set of equations of motion for all components of the conserved currents. Finally, we showed that the same equations of motion as derived in this paper can also be obtained from the 14-moment approximation. The reason is that, in the 14-moment approximation, the moments that involve high powers of $p_z/E_p$ are simply dropped, consistently with the solution at the free-streaming fixed point where the distribution function is peaked around $p_z=0$.

The present formulation of chiral hydrodynamics, just like conventional transient hydrodynamics, features free-streaming fixed points and attractor solutions, and therefore is applicable already at early time, i.e.,  before local equilibrium is reached. Furthermore, in the setup used in this paper, any parity-violating moments decay exponentially with $\tau/\tau_R$. This means that any initial \bbb{global} polarization of massless particles, which is not parallel to the fluid velocity, is unlikely to survive until freeze-out in a heavy-ion collision, unless the relaxation time $\tau_R$ is very large. One should however note that the assumption of translational invariance prohibits a nonzero fluid vorticity and is therefore only applicable to central heavy-ion collisions. On the other hand, it is known that in noncentral heavy-ion collisions, large values of the vorticity can be reached~\cite{STAR:2017ckg}. It will therefore be an interesting extension of this work to relax the assumption of translational invariance and describe a system with vorticity. It is expected that in this case, also parity-violating moments decay with power laws in the hydrodynamic regime, due to equilibrium contributions from the vorticity. Furthermore, while in this paper we considered massless particles with chiral degrees of freedom, it will be interesting to extend this study to massive particles with spin.

\section*{Acknowledgements}

We thank Jorge Noronha for enlightening discussions about chiral kinetic theory. N.W.\ acknowledges support by the German National Academy of Sciences Leopoldina through the Leopoldina fellowship program with funding code  LPDS 2022-11.

\begin{appendix}

\section{Coefficients of the moment equations}
\label{coeffapp}

The coefficients in \eq\eqref{dzdtau} read
\begin{align}
    a_{n\ell}&\equiv 1-n \left(\frac{(n+1-\ell)(n+1+\ell)}{(2n+1)(2n+3)}+\frac{(n+\ell)(n-\ell)}{(2n+1)(2n-1)}\right)+(n+\ell)\frac{n-\ell}{2n-1}\; ,\n\\
    b_{n\ell}&\equiv -n \frac{(n+\ell)(n-1+\ell)}{(2n+1)(2n-1)}+(n+\ell)\frac{n-1+\ell}{2n-1}\; ,\n\\
    c_{n\ell}&\equiv -n \frac{(n+1-\ell)(n+2-\ell)}{(2n+1)(2n+3)}\; .
\end{align}
 Furthermore, the coefficients in \eq\eqref{dxdtau} are given by
\begin{align}
    \bar{a}_{n\ell}&\equiv 1-(n-1) \left(\frac{(n+1-\ell)(n+1+\ell)}{(2n+1)(2n+3)}+\frac{(n+\ell)(n-\ell)}{(2n+1)(2n-1)}\right)+(n+\ell)\frac{n-\ell}{2n-1}\; ,\n\\
    \bar{b}_{n\ell}&\equiv -(n-1) \frac{(n+\ell)(n-1+\ell)}{(2n+1)(2n-1)}+(n+\ell)\frac{n-1+\ell}{2n-1}\; ,\n\\
    \bar{c}_{n\ell}&\equiv -(n-1) \frac{(n+1-\ell)(n+2-\ell)}{(2n+1)(2n+3)}\; .
\end{align}

\section{Analysis of exact free-streaming solution}
\label{freeapp}

Consider the free-streaming solution of \eq\eqref{boltzbjor}, obtained in the limit  $\tau_R\rightarrow\infty$,
\begin{equation}
 f^\lambda(\tau,p_\bot,p_z)= f^\lambda_\text{in}\left(\p_\bot,p_z\frac{\tau}{\tau_0}\right) ,\label{freesol}
\end{equation}
where $f^\lambda_\text{in}\left(\p_\bot,p_z\right)$ denotes the initial distribution. 
This explicit solution makes it possible to analyze the early- and late-time behaviors of the chiral moments by inserting \eq\eqref{freesol} into \eqs\eqref{momdefs}. We obtain
\begin{align}
 \J_{n\ell} \equiv \int d^3p\, Y_n^\ell(\theta,\phi) f^\lambda_\text{in}\left(\p_\bot,p_z\frac{\tau}{\tau_0}\right)
=\frac{\tau_0}{\tau} \int d^3p\, \mathcal{P}_n^\ell(p_z/\varepsilon_{p\tau}) e^{i\ell\phi} f^\lambda_\text{in}\left(\p_\bot,p_z\right)
\end{align}
where $\varepsilon_{p\tau}\equiv(\tau/\tau_0)\sqrt{(p_z\tau_0/\tau)^2+\p_\bot^2}$. 

It follows that,  for early times, $\tau\ll\tau_0$,
\begin{align}
    \J_{n\ell}&= \frac{\tau_0}{\tau} \int d^3p\, \left(\frac{p_\bot}{\epsilon_{p\tau}}\right)^\ell\frac{\mathcal{P}_n^\ell(p_z/\varepsilon_{p\tau})}{\left(\sqrt{1-(p_z/\varepsilon_{p\tau})^2}\right)^\ell} e^{i\ell\phi} f^\lambda_\text{in}\left(\p_\bot,p_z\right)\n\\
    &\rightarrow\left(\frac{\tau_0}{\tau}\right)^{1-\ell} \int d^3p\, \left[\text{sgn}(p_z)\right]^{n+\ell} \left(\frac{p_\bot}{|p_z|}\right)^\ell B_{n\ell}\, e^{i\ell\phi} f^\lambda_\text{in}\left(\p_\bot,p_z\right)\sim \tau^{\ell-1}\; , \qquad \qquad (\tau/\tau_0\rightarrow 0)
    \label{zearly}
\end{align}
where in the last line we used the relation $\Pc_n^\ell(-x)=(-1)^{n+\ell}\Pc_n^\ell(x)$. Note that the coefficient $B_{n\ell}$ defined in \eq\eqref{bnell} is a finite number, while in the limit $x\rightarrow 1$, $\Pc_n^\ell(x)/(1-x^2)^{k/2}$ vanishes for   $k<\ell$ and diverges for $k>\ell$.

For late times, $\tau\gg\tau_0$, we find, for $n+\ell$ even,
\begin{equation}
    \J_{n\ell}\rightarrow\frac{\tau_0}{\tau} \int d^3p\, \mathcal{P}^\ell_n(0) e^{i\ell\phi} f^\lambda_\text{in}\left(\p_\bot,p_z\right)\sim \frac{1}{\tau}\; , \qquad \qquad (\tau/\tau_0\rightarrow \infty\; ,\ n+\ell \text{ even})\; .
\end{equation}
For $n+\ell$ odd,  $\mathcal{P}^\ell_n(0)=0$, so that   we need to consider the next-to-leading order in a Taylor expansion. We get
\begin{equation}
    \J_{n\ell}\rightarrow\left(\frac{\tau_0}{\tau}\right)^2 \int d^3p\, (n+\ell)\mathcal{P}^\ell_{n-1}(0) e^{i\ell\phi}  \frac{p_z}{p_\bot} f^\lambda_\text{in}\left(\p_\bot,p_z\right)\sim \frac{1}{\tau^2}\; , \qquad \qquad (\tau/\tau_0\rightarrow \infty\; ,\ n+\ell \text{ odd})\; , \label{zlate}
\end{equation}
where we used $\left(\mathcal{P}_n^\ell\right)^\prime(0)=(n+\ell)\mathcal{P}^\ell_{n-1}(0)$. 

A similar analysis for the moments $\T_{n\ell}$ yields
\begin{align}
    \T_{n\ell}&=  \left(\frac{\tau_0}{\tau}\right)^2 \int d^3p\, \varepsilon_{p\tau} \mathcal{P}_n^\ell(p_z/\varepsilon_{p\tau}) e^{i\ell\phi} f^\lambda_\text{in}\left(\p_\bot,p_z\right)\; ,\\
    \T_{n\ell}&\rightarrow \left(\frac{\tau_0}{\tau}\right)^{2-\ell} \int d^3p\, \left[\text{sgn}(p_z)\right]^{n+\ell} \frac{p_\bot^\ell}{|p_z|^{\ell-1}} B_{n\ell}\, e^{i\ell\phi} f^\lambda_\text{in}\left(\p_\bot,p_z\right)\sim \tau^{\ell-2}\; , &(\tau/\tau_0\rightarrow 0)\; ,\\ 
    \T_{n\ell}&\rightarrow \frac{\tau_0}{\tau}\int d^3p\, p_\bot \mathcal{P}^\ell_n(0)\, e^{i\ell\phi} f^\lambda_\text{in}\left(\p_\bot,p_z\right) \sim \frac{1}{\tau}\; , & (\tau/\tau_0\rightarrow \infty\; ,\   n+\ell \text{ even} )\; , \label{xlate}\\
    \T_{n\ell}&\rightarrow \left(\frac{\tau_0}{\tau}\right)^2\int d^3p\, p_z (n+\ell)\mathcal{P}^\ell_{n-1}(0) e^{i\ell\phi}  f^\lambda_\text{in}(\p_\bot,p_z) \sim \frac{1}{\tau^2}\; , & (\tau/\tau_0\rightarrow \infty\; ,\ n+\ell \text{ odd} )\; .
    \end{align}

\section{Example of an exact solution for free streaming}

Consider the initial distribution function 
\begin{equation}
    f_\text{in}^\lambda= \left(1+\frac{p^x}{\Lambda}+\frac{p^y}{\Lambda}\right)e^{-\p^2/\Lambda^2}\; 
\end{equation}
with $\Lambda$ being a constant. This function is used as initial condition for the plots of Sect.~\ref{freesec}.  The corrsponding  free-streaming solution reads
\begin{equation}
    f^\lambda= \left(1+\frac{p^x}{\Lambda}+\frac{p^y}{\Lambda}\right)e^{-[p_\bot^2+(\tau/\tau_0)^2p_z^2]/\Lambda^2}\; .
\end{equation}
We obtain
\begin{align}
   \text{Re}\, \mJ_{11}^\lambda&= \text{Re} \int d^3p\, Y_1^1(\theta,\phi)  f^\lambda\n\\
   &= -\frac\pi2 \Lambda^3 \int_{-1}^1d\cos\theta\, \left(1-\cos^2\theta\right)\, \frac{1}{\{1+[(\tau/\tau_0)^2-1]\cos^2\theta\}^2}\; .
\end{align}
Analogously we find
\begin{align}
 \text{Re}\, \mJ_{31}^\lambda&= \text{Re} \int d^3p\, Y_3^1(\theta,\phi)  f^\lambda\n\\
    &= \frac32\frac\pi2 \Lambda^3 \int_{-1}^1d\cos\theta\, \left(1-5\cos^2\theta\right) \left(1-\cos^2\theta\right)\, \frac{1}{\{1+[(\tau/\tau_0)^2-1]\cos^2\theta\}^2}\; .
\end{align}
The remaining integrals can be done numerically. Note that the initial conditions imply
\begin{equation}
 \text{Re}\, \mJ_{11}^\lambda (\tau=\tau_0)=-\frac{2\pi}{3} \Lambda^3, 
\end{equation}
and, for $n=3,5,7,\ldots$,
\begin{equation}
    \text{Re}\, \mJ_{n1}^\lambda(\tau=\tau_0)=0.
\end{equation}

\section{14-moment approximation}
\label{14app}

In this appendix we examine the comparison between our approach and the 14-moment approximation which was outlined in Sect.~\ref{compsec}.
We consider the energy-momentum tensor and the charge current for massless particles in Bjorken symmetry with parity breaking in the transverse plane, allowing for nonzero spatial components of the charge current and nondiagonal components of the energy-momentum tensor. 

As an example, we will discuss the $x$-component of the charge current,
\begin{equation}
    J^x\equiv \int d^3p\, \frac{p_x}{E_p} f\equiv -\text{Re} \mathcal{J}_{11}\; ,
\end{equation}
and the diagonal components of $T^{\mu\nu}$,
\begin{align}
 T^{00}&=\int d^3p\, E_p f\equiv \mathcal{T}_{00}\; ,\n\\
  T^{zz}  &= \int d^3p\, \frac{p_z^2}{E_p} f \equiv  \left( \frac23 \mathcal{T}_{20}+\frac13\mathcal{T}_{00}\right)\;, \n\\ T^{xx}&=T^{yy}=\frac12(T^{00}-T^{zz})\equiv\sum_{\lambda=\pm1}\frac13(\mathcal{T}_{00}^\lambda-\mathcal{T}_{20}^\lambda) \; ,
  \end{align}
 with $\J_{n\ell}$ and $\T_{n\ell}$ given by \eq\eqref{momdefs}. Other components of the charge current and energy-momentum tensor can be treated analogously.
  We furthermore define the moments
  \begin{equation}
      \X_n\equiv \int d^3p\, \frac{p_x}{E_p} \left(\frac{p_z}{E_p}\right)^n f\; ,\qquad \qquad \Z_n\equiv \int d^3p\, E_p \left(\frac{p_z}{E_p}\right)^n f\; ,
      \label{ISmom}
  \end{equation}
  in terms of which, we have
  \begin{equation}
     J^x\equiv \X_0\; , \qquad \qquad T^{00}\equiv \Z_0\; , \qquad \qquad T^{zz}\equiv \Z_2\; .
  \end{equation}
  These moments are the ones originally used by Israel and Stewart~\cite{Israel:1979wp}. Their disadvantage is that they are not orthogonal. However, when used in the context of the free Bjorken expansion which suppresses higher values of $p_z/E_p$, their advantage is that all $\X_n$ and $\Z_n$ become smaller by a power $p_z/E_p$ at any order of $n$. This is not the case for $\mathcal{T}_{n\ell}$ and $\mathcal{J}_{n\ell}$.
  
  We first consider the equation of motion for $J^x$ using the moments \eqref{ISmom}, 
  \begin{equation}
      \partial_\tau \X_0=-\frac{1}{\tau} \left(  \X_0-\X_2 \right)-\frac{1}{\tau_R}\X_0\; .
  \end{equation}
  In the 14-moment approximation, we simply drop the term proportional to $\X_2$, since it does not appear in any conserved current. This is the analog of what we called the naive truncation.  For an expanding system with Bjorken symmetry, this is a good approximation, since the expansion naturally suppresses the higher orders of $p_z/E_p$ that are contained in $\X_2$.  Thus we obtain
  \begin{equation}
      \partial_\tau \X_0=-\frac{1}{\tau}   \X_0-\frac{1}{\tau_R}\X_0\; .
      \label{J1}
  \end{equation}
  On the other hand, when $\mathcal{J}_{11}$ is calculated with the orthogonal moments \eqref{momdefs}, one gets
  \begin{equation}
      \partial_\tau \mathcal{J}_{11}=-\frac{1}{\tau}\left(\frac45\mathcal{J}_{11}-\frac{2}{15}\mathcal{J}_{31}\right)-\frac{1}{\tau_R}\mathcal{J}_{11}\; .
  \end{equation}
  In this case, we should not simply drop $\mathcal{J}_{31}$, since it contains not only terms proportional to $(p_z/E_p)^3$, but also terms proportional to $p_z/E_p$. Instead, implementing the so-called fixed point truncation, we replace it by the value at the free-streaming fixed point, $\mathcal{J}_{31}\simeq-(3/2)\mathcal{J}_{11}$. We then arrive at
  \begin{equation}
      \partial_\tau \mathcal{J}_{11}=-\frac{1}{\tau} \mathcal{J}_{11}-\frac{1}{\tau_R}\mathcal{J}_{11}\; ,
  \end{equation}
  which is identical to \eq\eqref{J1}. The same reasoning holds for all the components of the charge current and the energy-momentum tensor which vanish in equilibrium and whose equations of motion do not couple to those of other components.
  
  On the other hand, for the diagonal components of the energy-momentum tensor the situation is a bit different, since they are nonzero in local equilibrium. Furthermore, the equations of motion are coupled and the truncation using fixed points is not unique. In order to obtain the diagonal components of the energy-momentum tensor, we need to determine two moments, either $\mathcal{T}_{00}$ and $\mathcal{T}_{20}$, or $\mathcal{Z}_0$ and $\mathcal{Z}_2$. These satisfy the following coupled equations of motion:
  \begin{align}
      \partial_\tau \mathcal{T}_{00}&= -\frac{1}{\tau}\frac23(2 \mathcal{T}_{00}+ \mathcal{T}_{20})\; ,\n\\
            \partial_\tau \mathcal{T}_{20}&= -\frac{1}{\tau}\left(\frac{38}{21} \mathcal{T}_{20}+\frac{8}{15} \mathcal{T}_{00}-\frac{12}{35}\mathcal{T}_{40}\right)-\frac{1}{\tau_R}\mathcal{T}_{20}\; , \label{L}
  \end{align}
  or
 \begin{align}
      \partial_\tau \mathcal{Z}_0&= -\frac{1}{\tau}(\mathcal{Z}_0+\mathcal{Z}_2)\; ,\n\\
            \partial_\tau \mathcal{Z}_2&= -\frac{1}{\tau}\left(3\mathcal{Z}_2-\mathcal{Z}_4\right)-\frac{1}{\tau_R}(\mathcal{Z}_2-\mathcal{Z}_{2,\text{eq}})\; , \label{T}
  \end{align}
  depending on the set of moments we use. 
  
  Now, in order to close the system of equations of motion in the fixed-point truncation, c.f.\ Ref.~~\cite{Blaizot:2021cdv}, we  may use
  \begin{equation}
      \mathcal{T}_{40}\simeq \Pc^0_{4}(0)\mathcal{T}_{00}=\frac38 \mathcal{T}_{00}\; 
      \label{L0L4}
  \end{equation}
  or
   \begin{equation}
      \mathcal{T}_{40}\simeq \Pc_{4}^0(0)/\Pc_2^0(0)\mathcal{T}_{20}=-\frac34 \mathcal{T}_{20}\; . \label{L2L4}
  \end{equation}
  We obtain
   \begin{align}
      \partial_\tau \mathcal{T}_{00}&= -\frac{1}{\tau}\frac23(2 \mathcal{T}_{00}+ \mathcal{T}_{20})\; ,\n\\
            \partial_\tau \mathcal{T}_{20}&= -\frac{1}{\tau}\left(\frac{38}{21} \mathcal{T}_{20}+\frac{17}{42} \mathcal{T}_{00}\right)-\frac{1}{\tau_R}\mathcal{T}_{20}\; , \label{Ltrun1}
  \end{align}
  or
  \begin{align}
      \partial_\tau \mathcal{T}_{00}&= -\frac{1}{\tau}\frac23(2 \mathcal{T}_{00}+ \mathcal{T}_{20})\; ,\n\\
            \partial_\tau \mathcal{T}_{20}&= -\frac{1}{\tau}\left(\frac{31}{15} \mathcal{T}_{20}+\frac{8}{15} \mathcal{T}_{00}\right)-\frac{1}{\tau_R}\mathcal{T}_{20}\; . \label{Ltrun2}
  \end{align}
  
  On the other hand, for the system of equations \eqref{T} we use the  14-moment approximation, neglecting the dissipative parts of all moments which are not components of the energy-momentum tensor, i.e.,
  \begin{equation}
      \mathcal{Z}_4\simeq \mathcal{Z}_{4,\text{eq}}\; . \label{T4}
  \end{equation}
  Thus \eqs\eqref{T} reduce to
  \begin{align}
      \partial_\tau \mathcal{Z}_0&= -\frac{1}{\tau}(\mathcal{Z}_0+\mathcal{Z}_2)\; ,\n\\
            \partial_\tau \mathcal{Z}_2&= -\frac{1}{\tau}(3\mathcal{Z}_2-\mathcal{Z}_{4,\text{eq}})-\frac{1}{\tau_R}(\mathcal{Z}_2-\mathcal{Z}_{2,\text{eq}})\; , \label{Ttrun1}
  \end{align}
  which translates into
  \begin{align}
      \partial_\tau \mathcal{T}_{00}&= -\frac{1}{\tau}\frac23(2 \mathcal{T}_{00}+ \mathcal{T}_{20})\; ,\n\\
            \partial_\tau \mathcal{T}_{20}&= -\frac{1}{\tau}\frac23\left(4 \mathcal{T}_{20}+ \frac54\mathcal{T}_{00}-\frac{3}{10}\mathcal{T}_{00,\text{eq}}\right)-\frac{1}{\tau_R}\mathcal{T}_{20}\; . \label{LtrunT1}
  \end{align}
  It is important to note that in the usual 14-moment approximation, the ideal part of $\mathcal{Z}_4$ is kept, and only the nonequilibrium part is set to zero. This is of course not a good approximation in the free-streaming regime, which  is far from local equilibrium. Instead, we should neglect the full $\mathcal{Z}_4$ at the free-streaming fixed point,
   \begin{equation}
      \mathcal{Z}_4\simeq 0\; . \label{T40}
  \end{equation}
  Then \eqs\eqref{T} become
  \begin{align}
      \partial_\tau \mathcal{Z}_0&= -\frac{1}{\tau}(\mathcal{Z}_0+\mathcal{Z}_2)\; ,\n\\
            \partial_\tau \mathcal{Z}_2&= -\frac{1}{\tau}3\mathcal{Z}_2-\frac{1}{\tau_R}(\mathcal{Z}_2-\mathcal{Z}_{2,\text{eq}})\; , \label{Ttrun2}
  \end{align}
  and therefore
  \begin{align}
      \partial_\tau \mathcal{T}_{00}&= -\frac{1}{\tau}\frac23(2 \mathcal{T}_{00}+ \mathcal{T}_{20})\; ,\n\\
            \partial_\tau \mathcal{T}_{20}&= -\frac{1}{\tau}\frac23\left(4 \mathcal{T}_{20}+ \frac54\mathcal{T}_{00}\right)-\frac{1}{\tau_R}\mathcal{T}_{20}\; . \label{LtrunT2}
  \end{align}

  While \eqs\eqref{Ltrun1}, \eqref{Ltrun2}, and \eqref{LtrunT2} agree at the free-streaming fixed point, \bbb{since then $\mathcal{T}_{20}=\Pc_2^0(0)\mathcal{T}_{00}=-(1/2)\mathcal{T}_{00}$,} and are expected to yield a good description of the dynamics in the free-streaming regime, they differ in their coefficients as soon as the system leaves the free-streaming point. On the other hand, \eq\eqref{LtrunT1} is not valid in the free-streaming regime. At the hydrodynamic fixed point, \eqs\eqref{L2L4} and \eqref{T4} remain valid, while \eqs\eqref{L0L4} and \eqref{T40} are not. Therefore, \eqs\eqref{Ltrun2} are expected to be a better approximation than \eqs\eqref{Ltrun1} or \eqref{LtrunT1} in the hydrodynamic regime. We conclude that, if one wants to use the 14-moment approximation to describe the evolution from free streaming to hydrodynamics, one should use some kind of interpolation between \eqs\eqref{T4} and \eqref{T40} in order to capture the dynamics of both regimes. This is similar to the approximation of $\mathcal{L}_4$ in the massive case~\cite{Jaiswal:2022udf}, where the moments $\mathcal{L}_n$ are not orthogonal anymore.
  
  We can also understand the differences between the different truncations by considering $\mathcal{T}_{40}$ itself,
  \begin{align}
    \mathcal{T}_{40} &=  \int d^3p\, E_p \frac18\left[35 \left(\frac{p_z}{E_p}\right)^4-30 \left(\frac{p_z}{E_p}\right)^2+3 \right]  f\; .
\end{align}
In the truncation \eqref{L0L4} both the first and the second term in the square brackets are dropped, while in the truncation \eqref{T40} only the first term is dropped. As long as the expansion governs the system, both approximations are reasonable since the distribution function is peaked around $p_z=0$. However, in the hydrodynamic regime this approximation is not justified, since the first term is nonzero in local equilibrium. In the truncation \eqref{T4}, the first term is approximated by its local-equilibrium value, which is only reasonable in the hydrodynamic regime. Finally, the truncation \eqref{L2L4} also keeps only the last term in the square brackets, but expresses this term through $\mathcal{T}_{20}$, which vanishes in local equilibrium and therefore in this case the issue in the hydrodynamic regime is avoided.

\end{appendix}

\bibliography{biblio_paper_long}{}

\end{document}